\documentclass{PoS}

\def\be{\begin{eqnarray}}
\def\ee{\end{eqnarray}}
\def\el{\nonumber\\}
\def\dslash{\makebox[0cm][l]{$\,/$}D}
\def\e{\mathrm{e}}
\def\tr{\mathrm{Tr}}
\def\conj#1{{\bar{#1}}}

\title{Partially quenched QCD with a chemical potential}

\ShortTitle{Partially quenched QCD with a chemical potential}

\author{\speaker{James C. Osborn}\\
        Physics Department \& Center for Computational Science\\
	Boston University\\
	Boston, MA 02215, USA\\
	E-mail: \email{josborn@physics.bu.edu}}

\abstract{
Using a chiral random matrix theory
we can now derive the low energy partition functions and
Dirac eigenvalue correlations of QCD with different chemical potentials for the
dynamical and valence quarks.
The results can also be extended to complex (and purely imaginary)
chemical potential.
We also discuss possible applications such as
fitting to low energy constants and understanding the phase diagrams of
the full and partially quenched theories.}

\FullConference{XXIVth International Symposium on Lattice Field Theory\\
                July 23-28, 2006\\
                Tucson, Arizona, USA}

\begin{document}

\section{Introduction}

Since the work of Banks and Casher \cite{BC} it has been known that the
low eigenvalues of the QCD Dirac operator are related to the breaking
of chiral symmetry.
Later Leutwyler and Smilga \cite{LS} showed that the low energy partition
 function for QCD was universal and could be determined directly from
 the finite volume chiral Lagrangian.
This led to a set of sum rules for inverse powers of the Dirac eigenvalues.
Shuryak and Verbaarschot \cite{ShV} then discovered that these sum rules could
also be obtained from a Random Matrix Theory (RMT) based on chiral symmetry.
Thus the low eigenvalue spectrum of the RMT is also universal and agrees
with QCD.
By now this agreement has been extensively checked by lattice simulations.

Since the RMT results contain an explicit volume dependence one use for
them has been to extract infinite volume results from finite volume
lattice simulations.
This was first done for the chiral condensate \cite{ccfit} since that
was the only available parameter in the RMT.
Since then it was realized that the low energy partition function of
quenched QCD with a chemical potential also depends on the pion decay
constant as does the eigenvalues \cite{TV}.
The determination of exact results for the low eigenvalue distributions
of quenched QCD with a chemical potential \cite{SpV} has then provided
one way to extract both constants in quenched lattice simulations \cite{OW}.
The eigenvalue distributions for unquenched QCD are also know \cite{O},
but due to the complex action they are not practical for current simulations.
One alternative that has been suggested is to use an imaginary isospin
chemical potential \cite{imu}.
Then the action is real and also a Hermitian eigensolver can still be used.
However for the unquenched simulations this required generating lattices
with an imaginary isospin chemical potential which, although it does not cause
any complications, it is not generally done.

The ideal case is to consider partially quenched ensembles where the dynamical
simulation is performed at one chemical potential (preferably zero) and the
eigenvalues are extracted using a different (nonzero) chemical potential.
We thus consider observables of the form
\be
\langle O \rangle = \frac{1}{Z} \int d[A]  ~O(\dslash(\mu_0))~
 \e^{-S_g(A)} \prod_{f=1}^{N_f} \det[\dslash(\mu_f)+m_f]
\ee
where $S_g$ is the gauge action and $\dslash(\mu)$ is the Dirac operator
with quark chemical potential $\mu$.
If the observable only depends on the low energy (zero momentum) modes
then the results are universal and can be evaluated using a chiral effective
theory such as RMT.
The observables we consider here are correlations of the Dirac
eigenvalues so that
\be
O(X) = \sum_k \delta^2(z_1-\lambda_k(X))
 ~\left\{ \sum_\ell \delta^2(z_2-\lambda_\ell(X)) \ldots \right\}
\ee
with $\lambda_k(X)$ the complex eigenvalues of $X$.
Partially quenched eigenvalue correlations with an imaginary isospin chemical
 potential have recently been published \cite{ADOS} and details of the
 calculation for a real chemical potential will appear elsewhere \cite{ADOS2}.
Here we present some details on the RMT used along with a sample of the results.

\section{Dirac eigenvalues from partially quenched partition functions}

One way to calculate correlations of Dirac eigenvalues is from the
partially quenched partition functions
\be
Z_{N_f,N_b}(\{m_f|m_b\},\{\mu_f|\mu_b\}) = \int d[A] \e^{-S_g(A)}
\frac{ \prod_{f=1}^{N_f} \det[\dslash(\mu_f)+m_f] }
     { \prod_{b=N_f+1}^{N_f+N_b} \det[\dslash(\mu_b)+m_b] } ~.
\ee
The partially quenched eigenvalue density for one dynamical flavor,
 for example, is given by \cite{rhoss}
\be
\rho_1(z,\mu_0|m_1,\mu_1) = \left[ \frac1\pi \partial_{\conj{z}} \partial_{z}
 Z_{3,2}(\{m_1,z,\conj{z}|m_b,\bar{m}_b\},\{\mu_1,\mu_0,-\mu_0|\mu_b,-\mu_b\})
 \right]_{m_b=z,\bar{m}_b=\bar{z},\mu_b=\mu_0} ~.
\ee
The presence of the extra (conjugate) flavors with negative $\mu$ is
required when the eigenvalue spectrum of the Dirac operator is complex.
The necessary partition functions can then in principle be obtained from
zero momentum chiral Lagrangians,
however these are only known for certain cases.
The purely fermionic partition function is given by \cite{TV}
\be
\label{chiL}
Z_{N_f,0} = \int_{U(N_f)} dU \det(U)^\nu \exp\left(
 \frac12 \Sigma V \tr\,M\{U+U^{\dagger}\} -\frac14 F^2 V \tr[U,Q][U^{\dagger},Q]
\right)
\ee
where $\Sigma$ and $F$ are the tree level condensate and pion decay constant
and $M$ and $Q$ are diagonal matrices of $m_f$ and $\mu_f$ respectively.
The bosonic partition function $Z_{0,2}$ is also known \cite{SpV}.
So far no direct evaluation of a partially quenched chiral Lagrangian
has been performed.
Instead the partially quenched partition functions can now be
 evaluated from RMT.

\section{RMT for QCD with a chemical potential}

Here we will consider a variant of the RMT introduced in \cite{O}.
The Dirac matrix in a chiral basis is written as (for zero mass)
\be
\label{dnew}
\mathcal{D}_f = \left( \begin{array}{cc}
0 & i a_f \Phi + b_f \Theta \\
i a_f \Phi^{\dagger} + b_f \Theta^{\dagger} & 0
\end{array} \right) ~.
\ee
Here $\Phi$ and $\Theta$ are complex $(N+\nu) \times N$ matrices
with $\nu$ the topological charge.
We start with a more general form of the Dirac matrix where
 $a_f \equiv a(T,\mu_f)$ and $b_f \equiv b(T,\mu_f)$ are functions of the
 temperature and chemical potentials that we will determine next.
The QCD partition function with $N_f$ quark flavors can now be modeled as
\be
\label{pfnew}
Z =
 \int d\Phi ~ d\Theta ~ \exp\left(-\alpha N \tr\left[ \Phi^\dagger \Phi + \Theta^\dagger \Theta\right]\right)
 \prod_{f=1}^{N_f} \det(\,\mathcal{D}_f + m_f\,) ~.
\ee
The constant $\alpha$ sets the average eigenvalue spacing and will also
 be determined next.
We could absorb it into $a$ and $b$, but instead will choose a normalization
 such that $a(T,\mu=0) = 1$ and $b(T,\mu=0) = 0$.

The coefficients can be determined by comparing to the chiral Lagrangian
at nonzero $\mu$.
The mapping follows very closely to that for $\mu=0$ given in \cite{HV}.
First the fermion determinants are all written in terms of Grassmann variables
as
\be
\det(\mathcal{D}_f + m_f) = \int d\bar\psi_{1,2}^f d\psi_{1,2}^f
\exp\left[
m_f \sum_k \bar\psi_k^f\psi_k^f
+ \bar\psi_1^f(i a_f \Phi + b_f \Theta)\psi_2^f
+ \bar\psi_2^f(i a_f \Phi^{\dagger} + b_f \Theta^{\dagger})\psi_1^f
\right] ~.
\ee
Then the Gaussian integration over the matrices $\Phi$ and $\Theta$ can be performed.
The result is
\be
Z \propto \int \prod_f d\bar\psi_{1,2}^f d\psi_{1,2}^f
\exp\left[
m_f \sum_k \bar\psi_k^f\psi_k^f
+ \sum_g \frac{a_f a_g - b_f b_g}{\alpha N}
 \bar\psi_1^f\psi_1^g \bar\psi_2^g\psi_2^f \right] ~.
\ee
If we consider only a baryon chemical potential such that all flavors
have the same chemical potential $\mu_f = \mu_B$, the coefficient of
the four fermion term becomes $[a^2(T,\mu_B) - b^2(T,\mu_B)]/\alpha N$.
At zero temperature we expect that the partition function does not depend
on the baryon chemical potential.
We can ensure this in the RMT by setting $a^2(0,\mu) - b^2(0,\mu) = 1$.
For nonzero temperature this will not be the case and one can leave the 
functions $a$ and $b$ independent.
As was mentioned previously \cite{lat04}, this could possibly be used to 
help map out the QCD phase diagram.

A nonlinear sigma model is obtained by
using a Hubbard-Stratonovitch transformation to break apart the four
fermion terms and then integrate out the Grassmann variables.
This gives
\be
Z = \int d\sigma \,\e^{-\alpha N \tr \sigma^\dagger \sigma}
 \det(\sigma+M)^{N+\nu}
 \det(\sigma^\dagger + P + M)^N
\ee
with $\sigma$ a $N_f\times N_f$ complex matrix and
$P = A\sigma^\dagger A - B\sigma^\dagger B - \sigma^\dagger$ such
that $P=0$ at $\mu_f=0$.
We also define $A$, $B$ and $M$ to be the diagonal matrices of $a_f$, $b_f$
 and $m_f$ respectively.
For small $m_f$ and $\mu_f$ the determinants can be expanded as
\be
 \det(\sigma+M)^{N+\nu} \det(\sigma^\dagger + P + M)^N &\approx&
 \exp\left(N\tr\left[
 \ln(\sigma^\dagger\sigma)+M\sigma^{-1}+(M+P)\sigma^{\dagger-1}
\right]\right) ~.
\ee
At $T=0$ we can also use the relation $A^2-B^2=1$ to expand $A$ as
$A \approx 1 + B^2/2$ which we will use below.
For $m_f=0$ and $\mu_f=0$ the large $N$ saddle point is given by
$\sigma^\dagger \sigma = 1/\alpha$.
Thus the Goldstone manifold is simply the unitary group \cite{HV}.
The low energy partition function is then
\be
\int dU \det(U)^{\nu} \exp\left( N \tr \sqrt{\alpha} M \{U+U^\dagger\}
  - N \tr [U,B][U^\dagger,B] \right) ~.
\ee
By matching terms with (\ref{chiL}) we find that
\be
\sqrt{\alpha} &\approx& \frac{\Sigma V}{2N} \el
b(T=0,\mu) &\approx& \mu F \sqrt{\frac{V}{2N}} \el
a(T=0,\mu) &\approx& 1 + \mu^2 F^2 \frac{V}{4N} ~.
\ee
With these definitions we then have a RMT that maps directly onto the
zero momentum chiral Lagrangian.
We note that the $O(\mu^2)$ term in $a$ above only contributes to the
overall normalization of the partition function.
We can therefore neglect it for most observables.
However quantities like the partition function, particle number
and related susceptibilities will depend on it.

\section{Partially quenched eigenvalue correlations}

A nice feature of the above RMT is that the partition function can be rewritten
directly in terms of the eigenvalues of the Dirac matrix (\ref{dnew}).
This was originally done for the unquenched model where the dynamical and
valence chemical potentials are all the same \cite{O}.
It is also possible to do this for the partially quenched case, although
the results become much more complicated.
Due to space constraints we will only show plots of some of the results and
save the details of the calculation for a future publication \cite{ADOS2}.

\begin{figure}[t]
  \vspace{-10mm}
  \begin{minipage}[t]{.5\textwidth}
    \begin{center}
      \includegraphics[angle=270,width=1.1\columnwidth,clip]{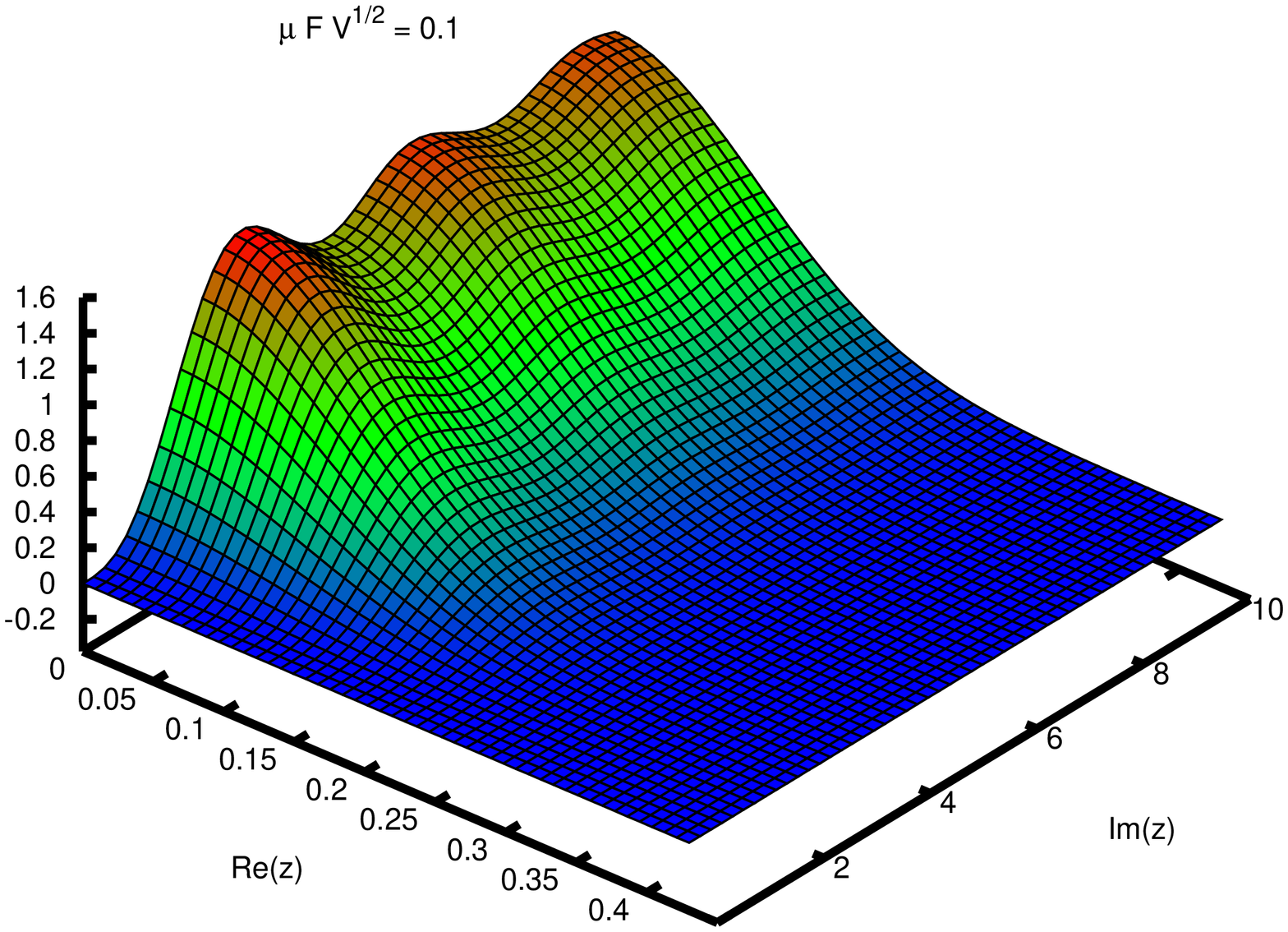}
    \end{center}
  \end{minipage}
  \vspace{-10mm}
  \begin{minipage}[t]{.5\textwidth}
    \begin{center}
      \includegraphics[angle=270,width=1.1\columnwidth,clip]{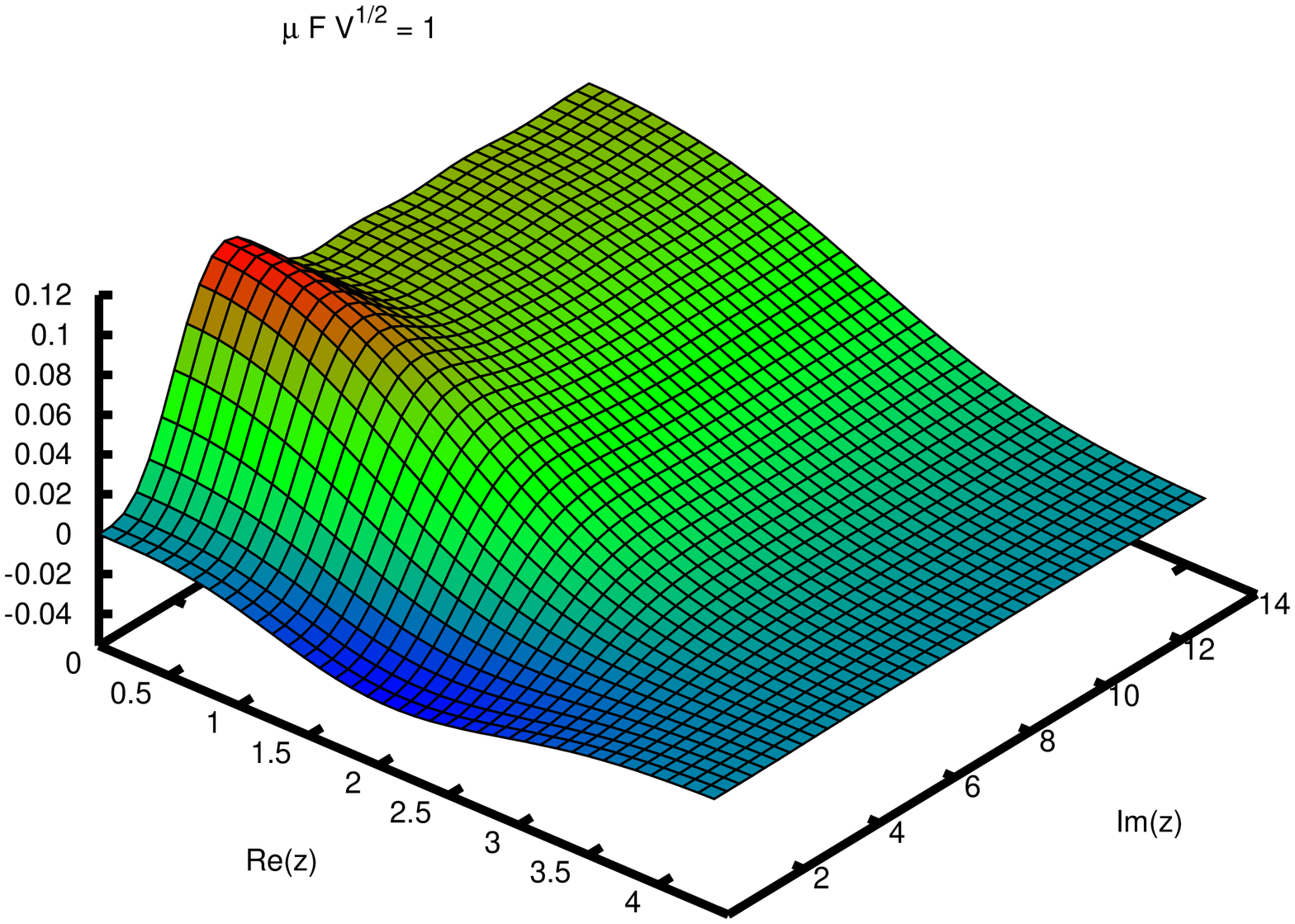}
    \end{center}
  \end{minipage}
  \hfill
  \begin{minipage}[t]{.5\textwidth}
    \begin{center}
      \includegraphics[angle=270,width=1.1\columnwidth,clip]{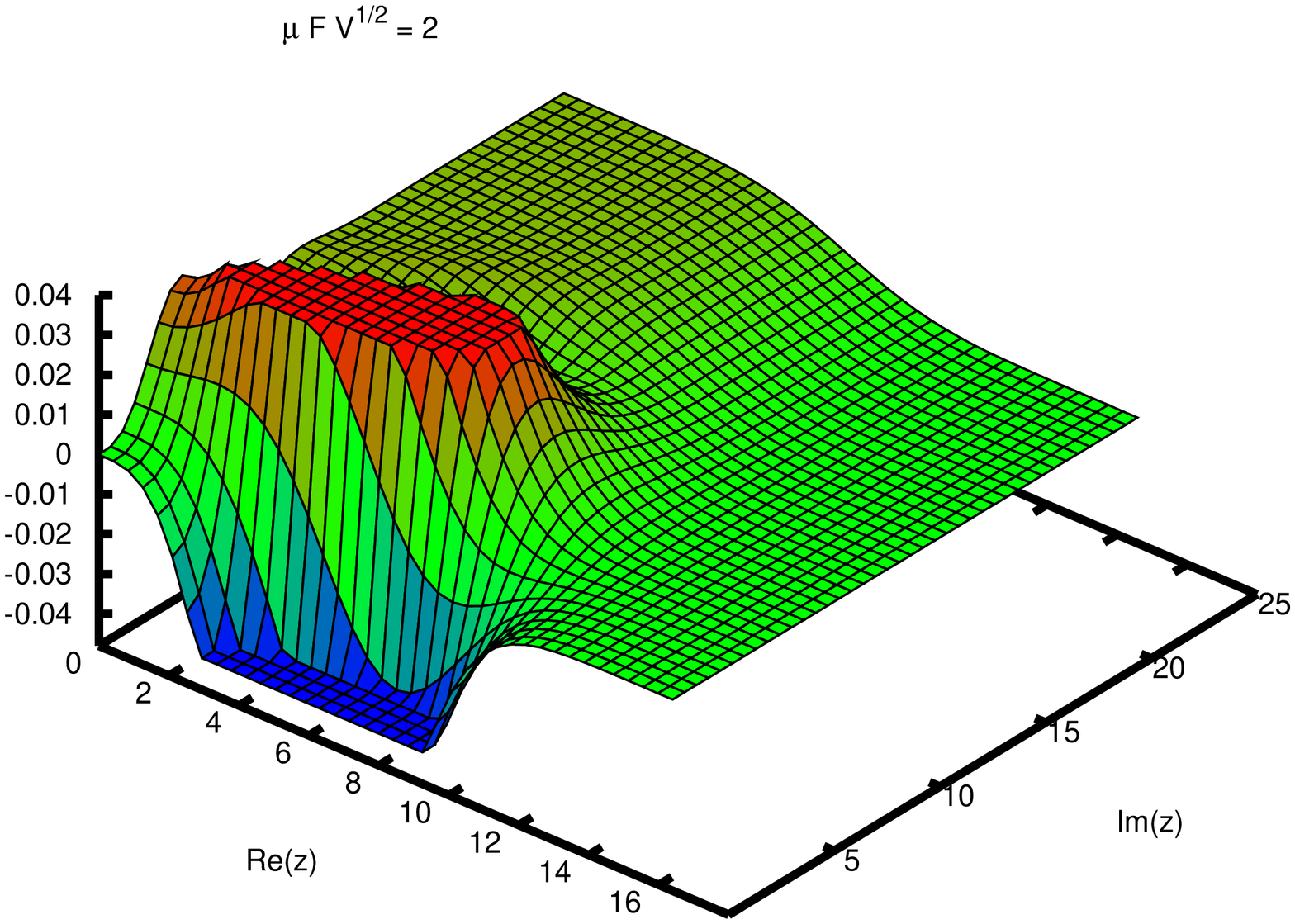}
    \end{center}
  \end{minipage}
  \begin{minipage}[t]{.5\textwidth}
    \begin{center}
      \includegraphics[angle=270,width=1.1\columnwidth,clip]{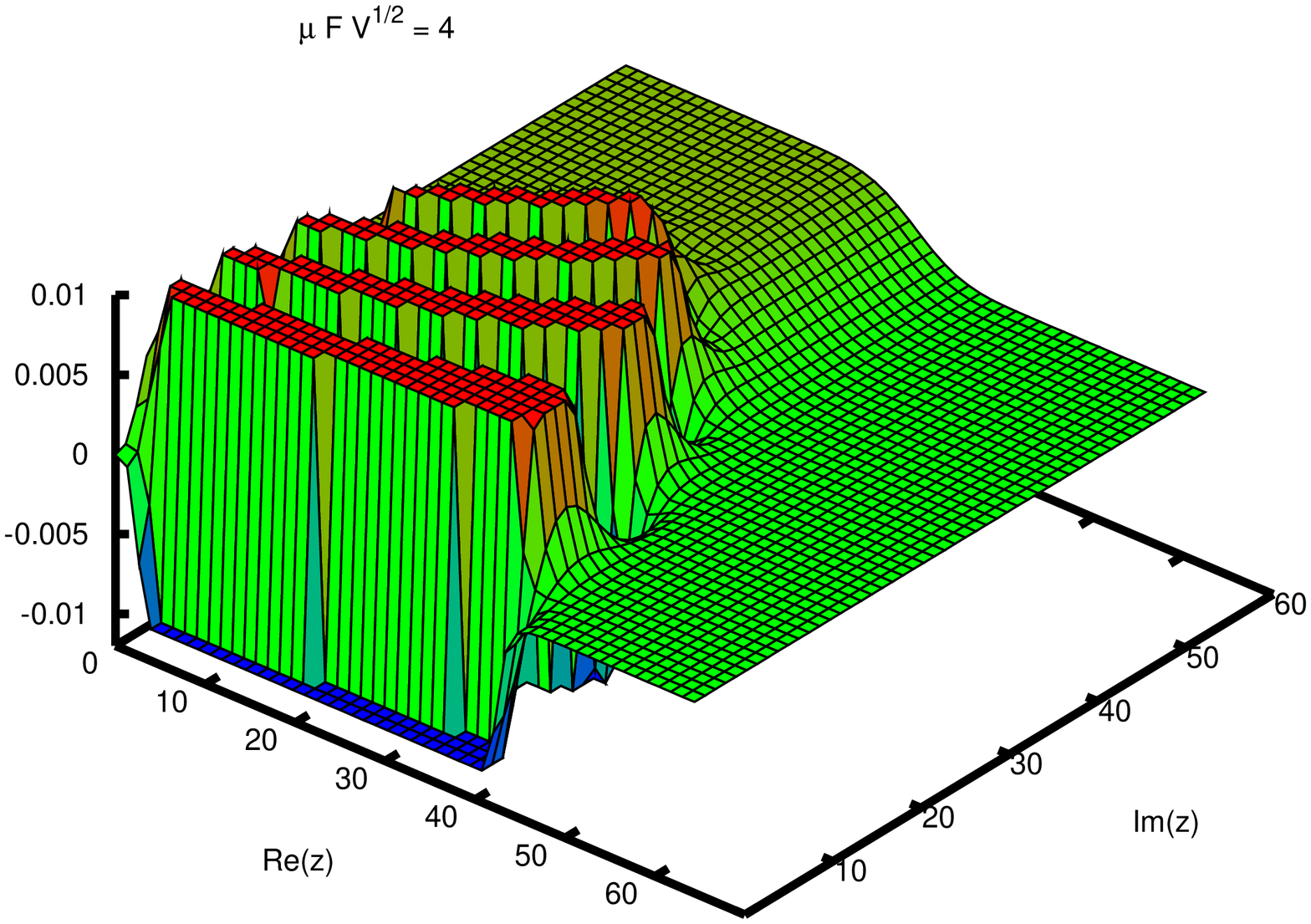}
    \end{center}
  \end{minipage}
  \hfill
  \vspace{-5mm}
  \caption{Real part of the one flavor eigenvalue density for dynamical
    mass $m=0$ at different chemical potentials $\mu$.
    The peaks have been clipped for better illustration at
    $\mu F \sqrt{V} = 2,4$.}
  \label{figd1}
\end{figure}

As a review we first show results for the one flavor eigenvalue density
 \cite{O,AOSV}.
In figure \ref{figd1} we plot the real part of the density for dynamical mass
$m=0$ for different values of the common chemical potential $\mu$.
At $\mu=0$ the eigenvalues are purely imaginary.
For small $\mu$ they begin to spread out into the complex plane, but still show
the characteristic oscillations along the imaginary axis due to chiral symmetry.
For larger $\mu$ those oscillations disappear and a new set of oscillations
appear around the real axis.
These oscillations grow exponentially with the volume and also change sign
unlike the ones for small $\mu$.
They are not seen in the quenched density and are in turn responsible for
producing a nonzero chiral condensate \cite{OSV}.

\begin{figure}[t]
  \vspace{-10mm}
  \begin{minipage}[t]{.5\textwidth}
    \begin{center}
      \includegraphics[angle=270,width=1.1\columnwidth,clip]{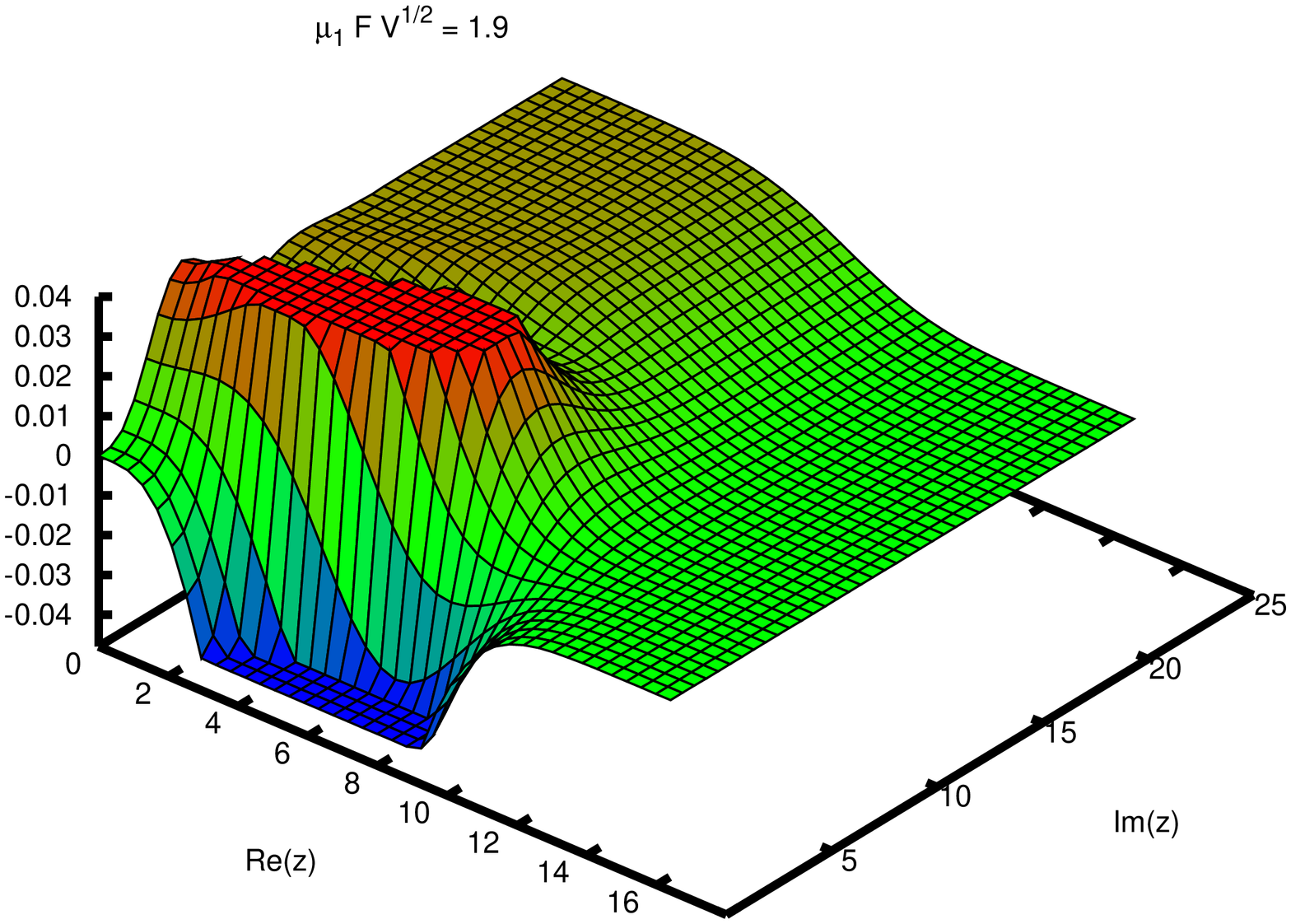}
    \end{center}
  \end{minipage}
  \vspace{-10mm}
  \begin{minipage}[t]{.5\textwidth}
    \begin{center}
      \includegraphics[angle=270,width=1.1\columnwidth,clip]{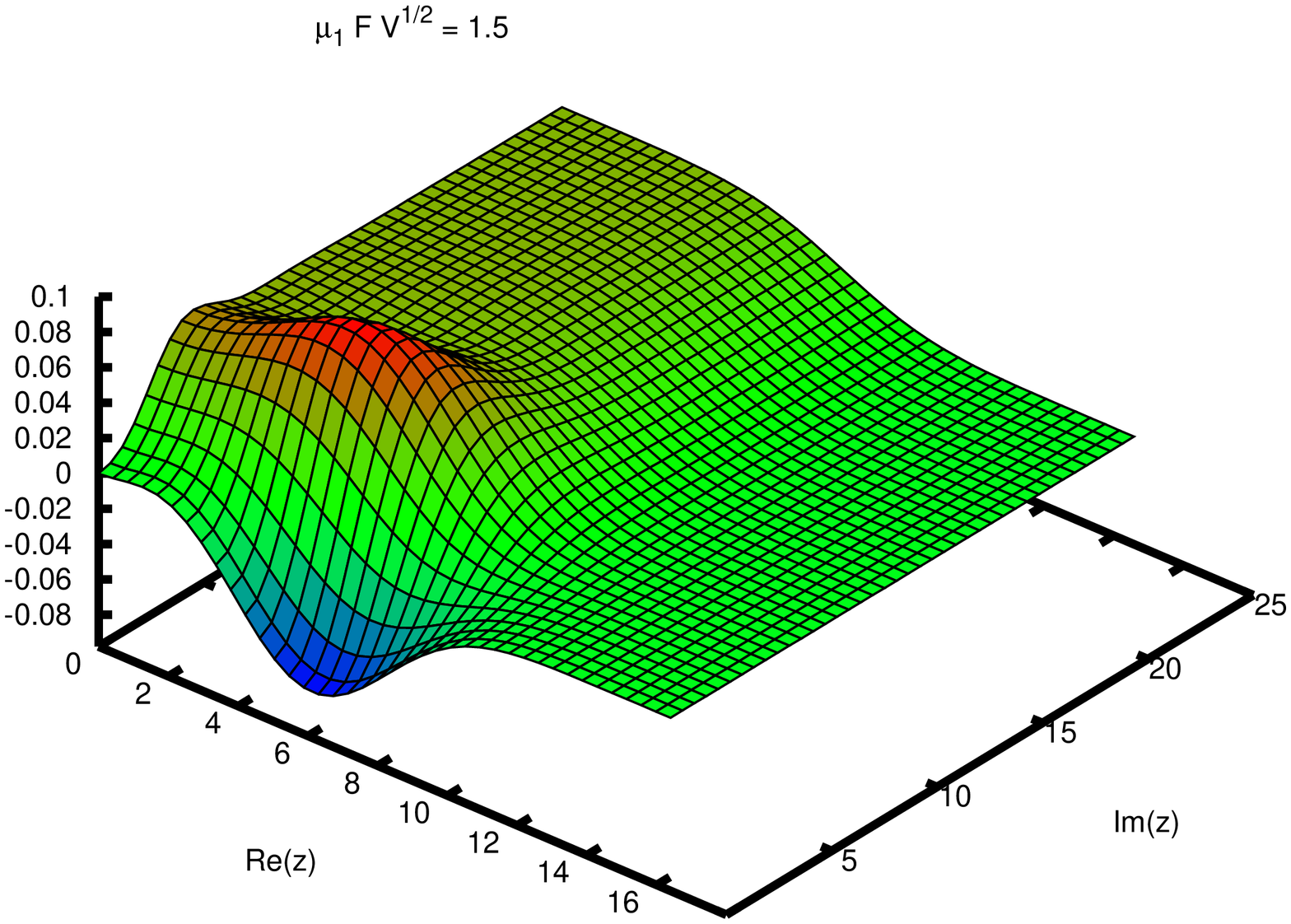}
    \end{center}
  \end{minipage}
  \hfill
  \begin{minipage}[t]{.5\textwidth}
    \begin{center}
      \includegraphics[angle=270,width=1.1\columnwidth,clip]{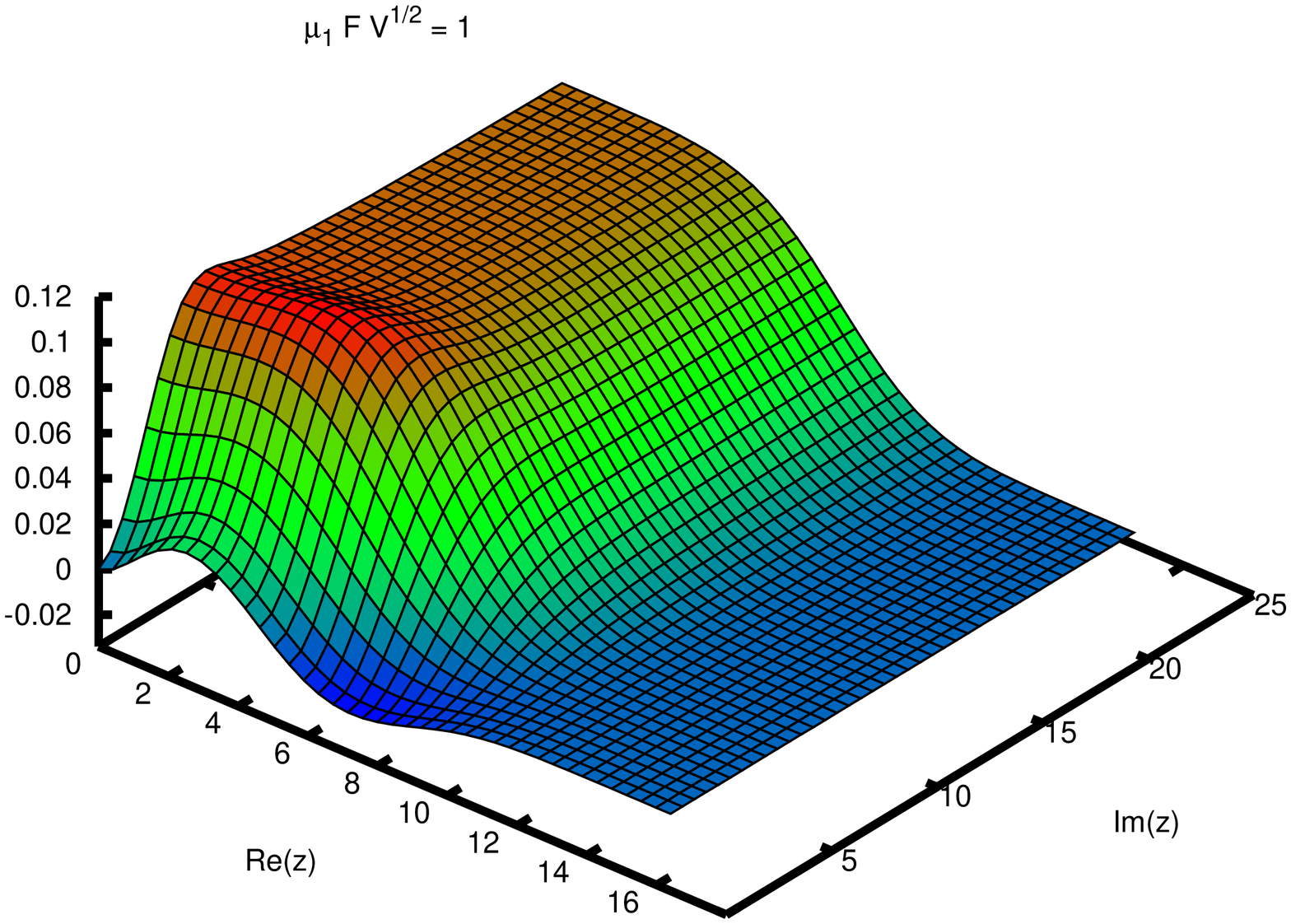}
    \end{center}
  \end{minipage}
  \begin{minipage}[t]{.5\textwidth}
    \begin{center}
      \includegraphics[angle=270,width=1.1\columnwidth,clip]{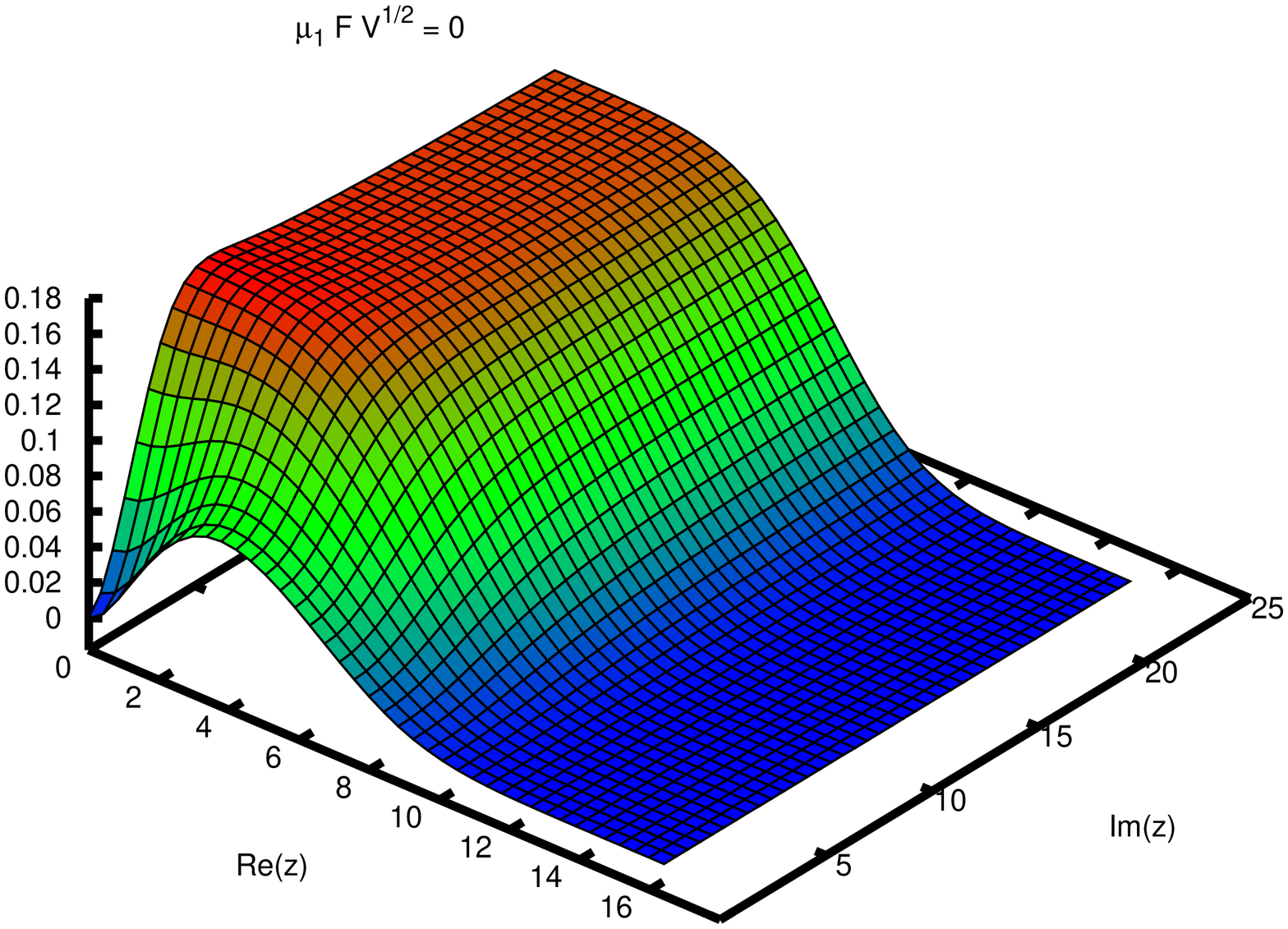}
    \end{center}
  \end{minipage}
  \hfill
  \vspace{-5mm}
  \caption{Real part of the partially quenched eigenvalue density for
    valence chemical potential $\mu_0 F \sqrt{V} = 2$ with dynamical mass
    $m_1=0$ and different dynamical chemical potentials $\mu_1$.
    The peak has been clipped for better illustration at
    $\mu_1 F \sqrt{V} = 1.9$.}
  \label{figdpq}
\end{figure}

\begin{figure}[t]
  \vspace{-4mm}
  \begin{minipage}[t]{.5\textwidth}
    \includegraphics[angle=270,width=1.1\columnwidth,clip]{dp0m0u2.eps}
  \end{minipage}
  \begin{minipage}[t]{.5\textwidth}
    \includegraphics[angle=270,width=1.1\columnwidth,clip]{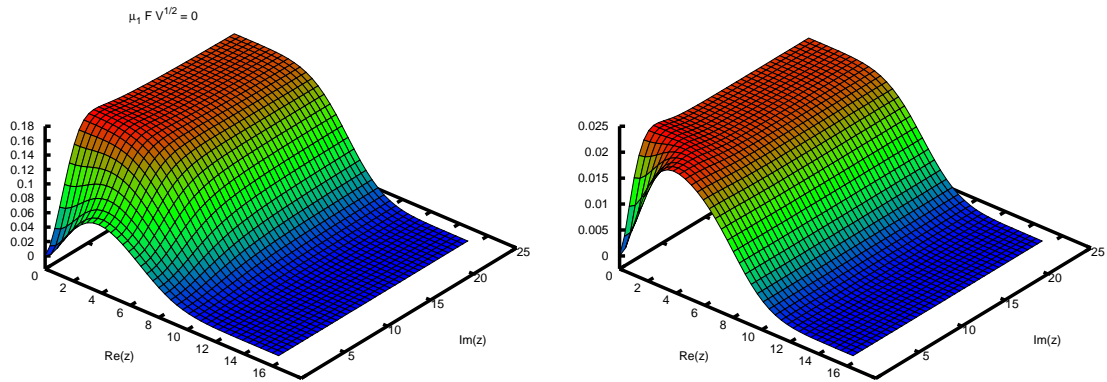}
  \end{minipage}
  \caption{Partially quenched (left) and quenched (right) eigenvalue density for
    valence chemical potential $\mu_0 F \sqrt{V}=2$.
    The dynamical chemical potential for the partially quenched is $\mu_1=0$
    with dynamical mass $m_1=0$.  Note the different scales.}
  \label{figdq}
\end{figure}

We now start with the one flavor density at $\mu F \sqrt{V} = 2$ and
lower the dynamical chemical potential $\mu_1$ as the valence chemical
is held fixed at $\mu_0 = \mu$.
This is show in figure \ref{figdpq}.
The peak that is seen near the real axis (which is clipped at
 $\mu_1 F \sqrt{V}= 1.9$) stays at about the same height as $\mu_1$ varies.
However the background density increases filling in the valley along
the real axis until the oscillations are completely covered at $\mu_1=0$.
At this point the density looks more like the quenched density shown
in figure \ref{figdq} than the one flavor result.
The main differences between the quenched and partially quenched in this
case are the overall scale and the relative size of the ``bump'' along
the real axis.

We can also extend these results to a complex chemical potential including
 a purely imaginary one.
This could be used to test analytic continuation of quantities from
 imaginary to real chemical potentials.
An imaginary chemical potential can also be used for fitting the
 low energy constants $\Sigma$ and $F$.
However in this case it is better to consider the mixed eigenvalue
 correlation between a valence and dynamical eigenvalue \cite{imu,ADOS},
 since it is much more sensitive to the chemical potential.

\section{Conclusions}

We have shown results for the eigenvalue correlations of partially
 quenched QCD with different dynamical and valence chemical potentials.
This could be useful for fitting low energy constants on existing lattices
 generated at zero chemical potential.
It is also possible to generate the low energy partition functions
 with different chemical potentials for each flavor from these results.
This could then be applied, for example, to study three flavor QCD.
Additionally the extension of the RMT to include coefficients which are
 general functions of temperature and chemical potential may be useful
 for mapping out the QCD phase diagram.

\end{document}